\documentclass{PoS}

\usepackage{mathtools}
\DeclareMathAlphabet{\mathcal}{OMS}{cmsy}{m}{n}

\newcommand{\pt}{\ensuremath{p_{\mathrm{T}}}}

\newcommand{\cmsSymbolFace}{\mathrm}

\newcommand{\TeV}{\ensuremath{\,\text{Te\hspace{-.08em}V}}}

\newcommand{\cPZ}{\ensuremath{\cmsSymbolFace{Z}}} 
\newcommand{\cPW}{\ensuremath{\cmsSymbolFace{W}}} 
\newcommand{\cPV}{\ensuremath{\cmsSymbolFace{V}}}
\newcommand{\cPqt}{\ensuremath{\cmsSymbolFace{t}}} 
\newcommand{\cPqb}{\ensuremath{\cmsSymbolFace{b}}} 
\newcommand{\cPaqt}{\ensuremath{\overline{\cmsSymbolFace{t}}}} 
\newcommand{\cPaqb}{\ensuremath{\overline{\cmsSymbolFace{b}}}} 
\newcommand{\bbbar}{\ensuremath{\cPqb\cPaqb}}
\newcommand{\ttbar}{\ensuremath{\cPqt\cPaqt}}

\newcommand{\cPH}{\ensuremath{\cmsSymbolFace{H}}} 

\title{Searches for ${\ttbar}\cPH$ production at CMS}

\ShortTitle{Searches for ${\ttbar}\cPH$ production at CMS}

\author{\speaker{Aruna Kumar Nayak}\thanks{On behalf of the CMS Collaboration}\\
        Institute of Physics, Bhubaneswar, India\\
        E-mail: \email{Aruna.Nayak@cern.ch}}

\abstract{Results of searches for the Higgs boson production in association with a pair of top quarks using the proton-proton collision data collected by the CMS detector at LHC are presented. First, results of the searches at centre-of-mass energy of 13 $\TeV$ are briefly discussed. Then, the combination of results at centre-of-mass energies of 7, 8, and 13 $\TeV$ are presented, which provides an observation of the Higgs boson production in association with a pair of top quarks with a significance of 5.2 standard deviations.}

\FullConference{XXVI International Workshop on Deep-Inelastic Scattering and Related Subjects (DIS2018)\\
		16-20 April 2018\\
		Kobe, Japan}

\begin{document}

\section{Introduction}
\begin{figure}
\begin{minipage}{0.5\linewidth}
\centerline{\includegraphics[width=0.5\linewidth]{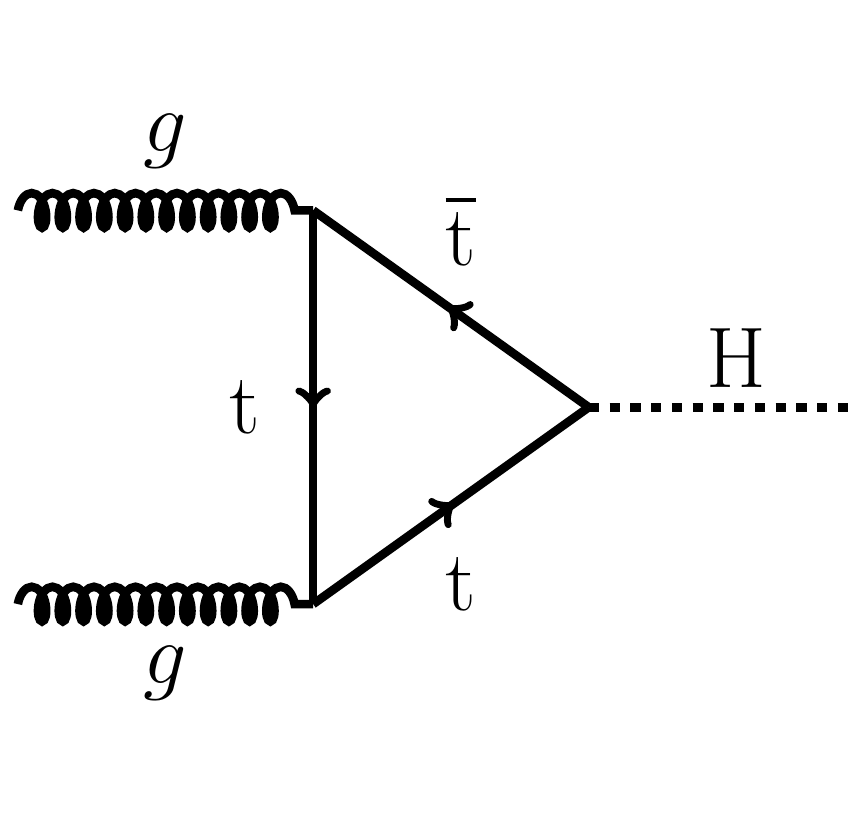}}
\end{minipage}
\hfill
\begin{minipage}{0.5\linewidth}
\centerline{\includegraphics[width=0.5\linewidth]{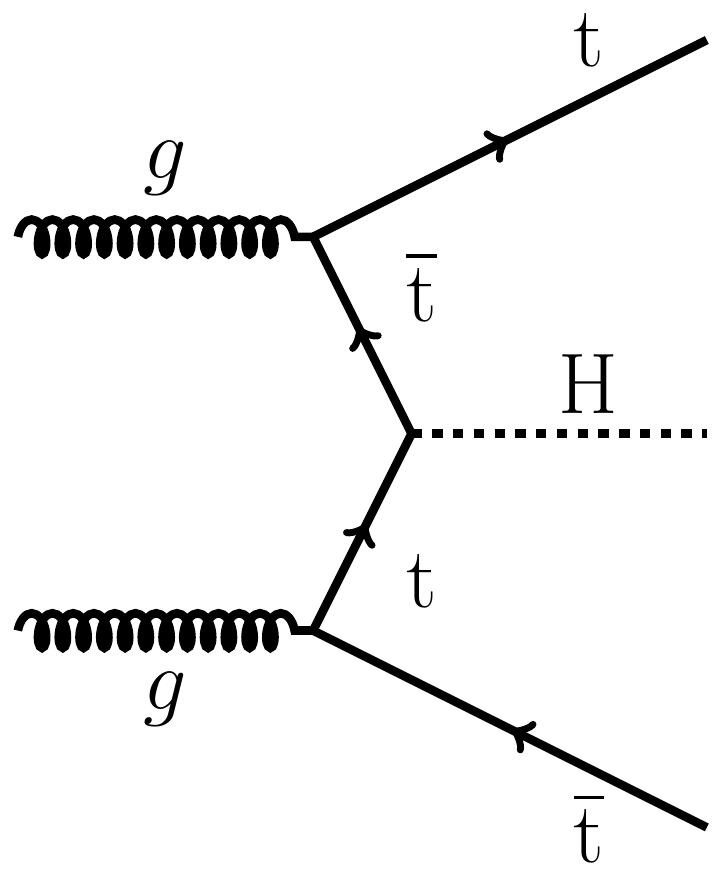}}
\end{minipage}
\caption[]{Feynman diagrams for the Higgs boson production via the gluon fusion process (left) and in association with a pair of top quarks (right).}
\label{fig:tth_diagram}
\end{figure}

Since the discovery of a Higgs boson by the ATLAS and CMS experiments in 2012~\cite{atlas_obs,cms_obs}, the major physics activity of these experiments at LHC has been centered on the precision measurement of properties of this particle to establish its exact nature.  
Towards this effort, the measurement of its coupling to the top quark has gained high phenomenological interest due to an extraordinary large value of the top quark mass compared to all other known particles. 
Though the top quark Yukawa coupling ($y_{\mathrm{t}}$) is measured indirectly from the Higgs boson production in the gluon fusion process, as shown in Fig.~\ref{fig:tth_diagram} (left), and agrees with the standard model (SM) expectation~\cite{HIG-15-002, CMS-PAS-HIG-17-031}, it can be affected by possible contributions of beyond-the-SM particles to the loop diagram.  
Hence, the measurement of the Higgs boson production in association with a top quark pair, ${\ttbar}\cPH$, provides the most precise model-independent estimate of $y_{\mathrm{t}}$. The diagram for the ${\ttbar}\cPH$ process is shown in Fig.~\ref{fig:tth_diagram} (right). 

In this article, we briefly discuss the search strategies and results of the ${\ttbar}\cPH$ production in various final states based on the proton-proton collisions recorded at centre-of-mass (CM) energy $\sqrt{s}$ = 13 $\TeV$ by the CMS detector~\cite{JINST:3:S08004} during 2016. 
Results of these searches combined with the same at $\sqrt{s}$ = 7 and 8 $\TeV$ are then presented. 

\section{Search for ${\ttbar}\cPH$, $\cPH~\to~\bbbar$}
This search targets for the final states where the Higgs boson decays to a pair of $\cPqb$ quarks~\cite{CMS-HIG-17-026, CMS-HIG-17-022}. Given the larger branching fraction of $\cPH~\to\bbbar$, this channel has higher statistics compared to other decays. However, it also suffers from copious contribution of backgrounds. The analyses are separately performed in the final states where the $\cPW$ boson, arising from the decay of a top quark, decays either to a lepton and a neutrino or to a pair of quarks. 

\subsection{${\ttbar}\cPH$, $\cPH~\to~\bbbar$ in semileptonic and leptonic final states}
This analysis is further divided into two separate searches: (1) where one $\cPW$ boson decays to a lepton and a neutrino and the other to a pair of quarks, and is named as the $\ell$+jets final state, and (2) where both the $\cPW$ bosons decay to leptons, and is called the dilepton final state. 


Both the event categories suffer from large contaminations from ${\ttbar}$+jets, especially from ${\ttbar}{\bbbar}$. Thus, discriminants based on multivariate methods, such as Boosted Decision Tree (BDT), Deep Neural Network (DNN), and matrix-element method (MEM) are used to separate signal from backgrounds~\cite{CMS-HIG-17-026}. 

The dilepton events are required to have at least two electrons, two muons, or one electron and one muon, and at least four jets. The events are further categorised depending on the number of $\cPqb$-tagged jets: $=$ 3$\cPqb$ or $\ge$ 4$\cPqb$. 
A BDT is trained for each of these two categories using input variables such as object kinematics, event shapes, and b-tag discriminants.
In the 3$\cPqb$ category the BDT distribution is used as the final discriminant to extract signal, while the 4$\cPqb$ category is further divided into two sub-categories based on the BDT distribution.
In both the sub-categories an MEM discriminant is used to distinguish signal from backgrounds, where the MEM discriminant is constructed from leading-order matrix elements for the ${\ttbar}\cPH$ and ${\ttbar}{\bbbar}$ processes. 


The $\ell$+jets events are required to have either an electron or a muon, $\ge$ 4 jets, and $\ge$ 3$\cPqb$-tagged jets. 
The events are divided into three categories based on the number of jets: 4, 5, and $\ge$ 6 jets. 
A multi-classification DNN is used in each of these jet multiplicity categories to discriminate signal from backgrounds. 
A DNN associates a set of probabilities describing the likelihood of an event being either signal ${\ttbar}\cPH$ or one of the ${\ttbar}$+jets backgrounds. 
Thus, events in each category are further divided to six sub-categories based on the most probable processes assigned by the DNN, forming a total of 18 jet-process sub-categories. 
In each sub-category, the DNN output distribution of the node that matches the process category is used as the final discriminant.


The discriminant distribution from all 18 sub-categories are fitted together in a maximum likelihood (ML) fit to extract the signal yield. 
The best-fit value of the signal strength modifier, which is the ratio of the signal production rate to that expected from the SM ($\mu~=~\sigma/\sigma_{SM}$), is shown in Fig~\ref{fig:tth_bb_results} (left), along with the uncertainties, separately for $\ell$+jets and dilepton final states as well as for their combination. Major sources of systematic uncertainties are the predictions for ${\ttbar}$ + heavy-flavour background rates, jet energy scale, and uncertainties on $\cPqb$-tagging efficiency and misidentification probabilities.
The significance of the signal is found to be 1.6 standard deviations ($\sigma$) compared to an expected significance of 2.2$\sigma$.


\subsection{${\ttbar}\cPH$, $\cPH~\to~\bbbar$ in fully hadronic final states}
Events where both the $\cPW$ bosons decay to quarks are considered in this study~\cite{CMS-HIG-17-022}. 
Given the final state comprises jets only, a special trigger, requiring at least six jets, large $H_{\mathrm{T}}$, where $H_{\mathrm{T}}$ is the sum $\pt$ of jets, and at least one or two $\cPqb$-tagged jets, is deployed to select events at the online level. 
The final state suffers from large QCD multijets in addition to ${\ttbar}$+jets and ${\ttbar}{\bbbar}$ backgrounds.
Since the multijet processes are dominated by gluon jets a quark--gluon likelihood discriminant is constructed to suppress this background. 
The events are divided into three categories based on the number of jets:  7, 8, and $\ge$ 9 jets. 
Each jet category is further divided to two sub-categories depending on the number of $\cPqb$-tagged jets, 3$\cPqb$ or $\ge$ 4$\cPqb$, resulting in a total of six sub-categories. 
The contribution and shape of multijet background in the signal region are obtained from a QCD enriched control region. 
The distributions of MEM discriminants, constructed from leading-order matrix elements for the ${\ttbar}\cPH$ and ${\ttbar}{\bbbar}$ processses, are used to extract signal. 

\begin{figure}
\begin{minipage}{0.5\linewidth}                                                                                                                                                                         
\centerline{\includegraphics[width=0.9\linewidth]{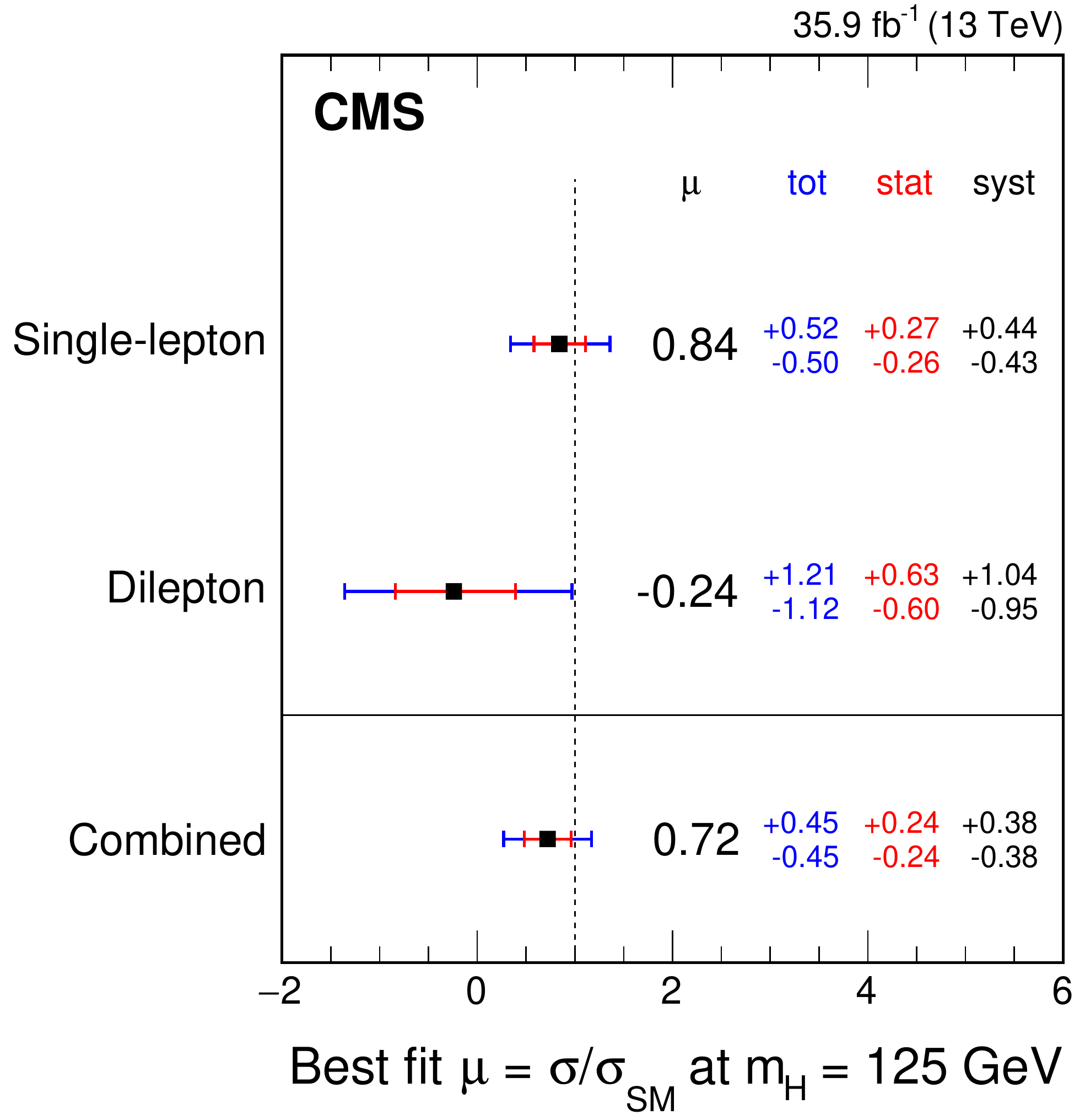}}                                                                                                                     
\end{minipage}
\hfill
\begin{minipage}{0.5\linewidth}
\centerline{\includegraphics[width=0.9\linewidth]{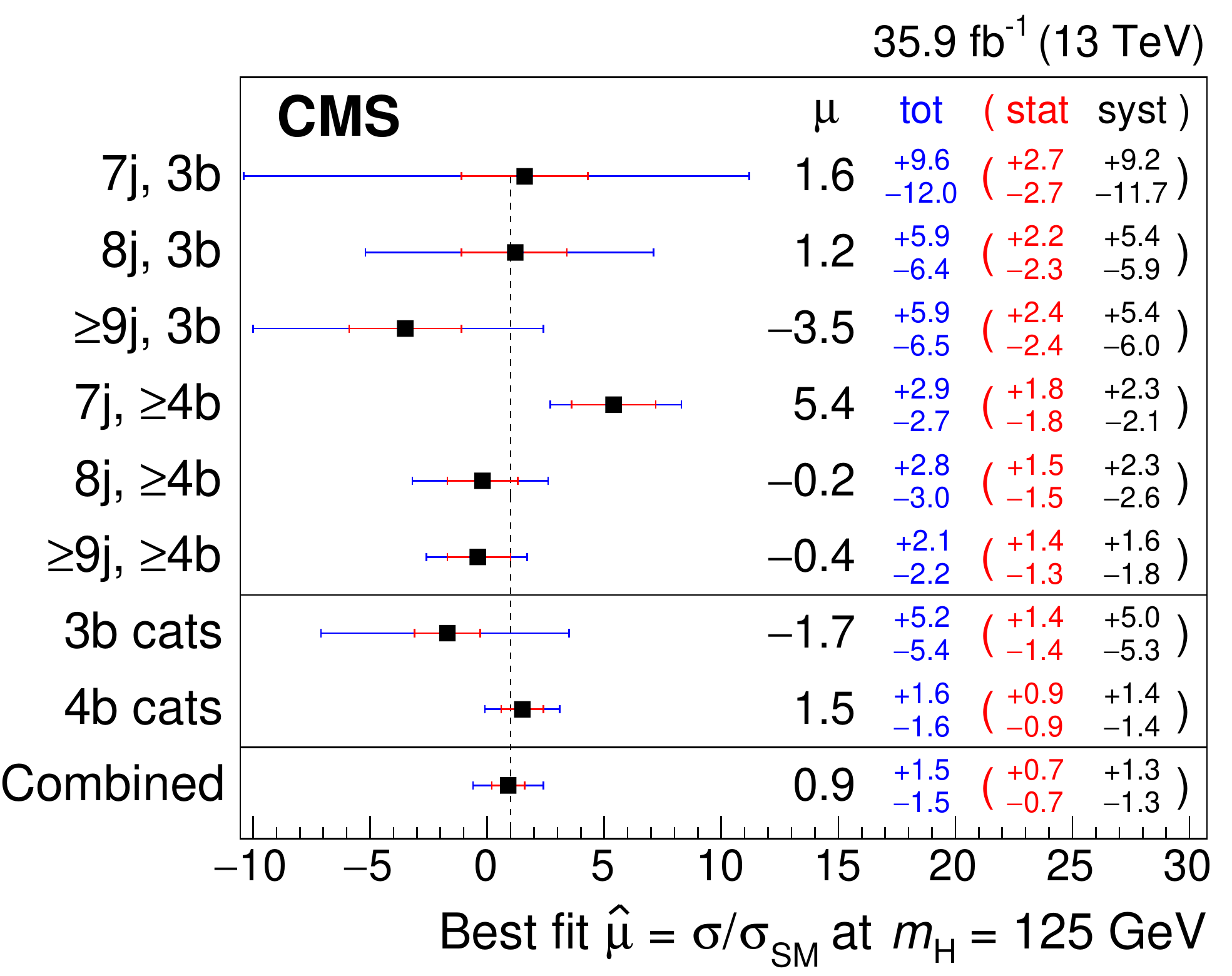}}
\end{minipage}
\caption[]{Best fit values of the signal strength modifiers with their 68\% expected confidence intervals (outer error bar), also split into their statistical (inner error bar) and systematic components, for the ${\ttbar}\cPH$, $\cPH~\to~\bbbar$ final states. (Left) For final states with at least one $\cPW$ boson decaying to leptons~\cite{CMS-HIG-17-026}. (Right) For final states with all $\cPW$ bosons decaying to quarks~\cite{CMS-HIG-17-022}.}
\label{fig:tth_bb_results}
\end{figure}

The best fit values of $\mu$ for each category and their combination are shown in Fig.~\ref{fig:tth_bb_results} (right), which are consistent with SM expectations.
The systematic uncertainties are dominated by the uncertainties on the estimation of multijet background, prediction of ${\ttbar}{\bbbar}$ background, $\cPqb$-tagging, and jet energy scale.  

\section{Search for ${\ttbar}\cPH$, $\cPH~\to~$leptons}
This final state corresponds to events where the Higgs boson decays to a pair of $\cPW$ bosons ($\cPH~\to~\cPW\cPW$), $\cPZ$ bosons ($\cPH~\to~\cPZ\cPZ$), or $\tau$ leptons ($\cPH~\to~\tau\tau$), with the subsequent decays of $\cPW$/$\cPZ$ bosons to electrons, muons, or tau leptons~\cite{CMS-HIG-17-018}. 


The events are categorized based on the number of leptons and hadronic $\tau$ decay candidates ($\tau_{h}$). 
The following six categories are considered: (1) one electron or muon and two $\tau_{h}$ (1$\ell$+2$\tau_{h}$), where $\ell$ and $\tau_{h}$ have opposite charge, (2) two leptons of opposite charge and no $\tau_{h}$ (2$\ell$ss), (3) two leptons of opposite charge and one $\tau_{h}$ (2$\ell$ss+1$\tau_{h}$), (4) three leptons and no $\tau_{h}$ (3$\ell$), where the sum of the charges of 3 leptons is 1, (5) three leptons and one $\tau_{h}$ (3$\ell$+1$\tau_{h}$), where the sum of the charges of four candidates is 0, (6) Four leptons (4$\ell$), where $\ell$ represents electron or muon. 
Additional selection criteria, such as requiring at least two to four jets, of which one to two $\cPqb$-tagged, threshold on missing transverse energy, and veto around the $\cPZ$ boson mass, are applied depending on event categories. 

SM processes namely ${\ttbar}\cPV$ ($\cPV$ = $\cPW$ and $\cPZ$) and diboson ($\cPW\cPW$, $\cPZ\cPZ$, and $\cPW\cPZ$) constitute the major irreducible backgrounds, while ${\ttbar}$+jets, where jets are misidentified as leptons, is also a major background. 
The irreducible backgrounds are modeled with Monte Carlo simulation and validated using data in dedicated control regions. 
Backgrounds with misidentified electron, muon, or $\tau_{h}$ candidates are estimated from data using a probabilistic method, where the jet $\to$ $\ell/\tau_{h}$ misidentification probabilities are measured from data in control regions as a function of $\pt$ and $\eta$. 

MEM and BDT discriminants are used to extract the signal yield. In some categories, such as 2$\ell$ss, 3$\ell$ and 3$\ell$+1$\tau_{h}$, two separate BDT discriminants are used, one trained against the ${\ttbar}\cPV$ background while the other against ${\ttbar}$+jets. The outputs of the two BDTs are mapped to a single discriminant using a likelihood method. In the 3$\ell$ category the MEM discriminant is also used as input to the BDT. 

\begin{figure}
\begin{minipage}{0.5\linewidth}
\centerline{\includegraphics[width=0.9\linewidth]{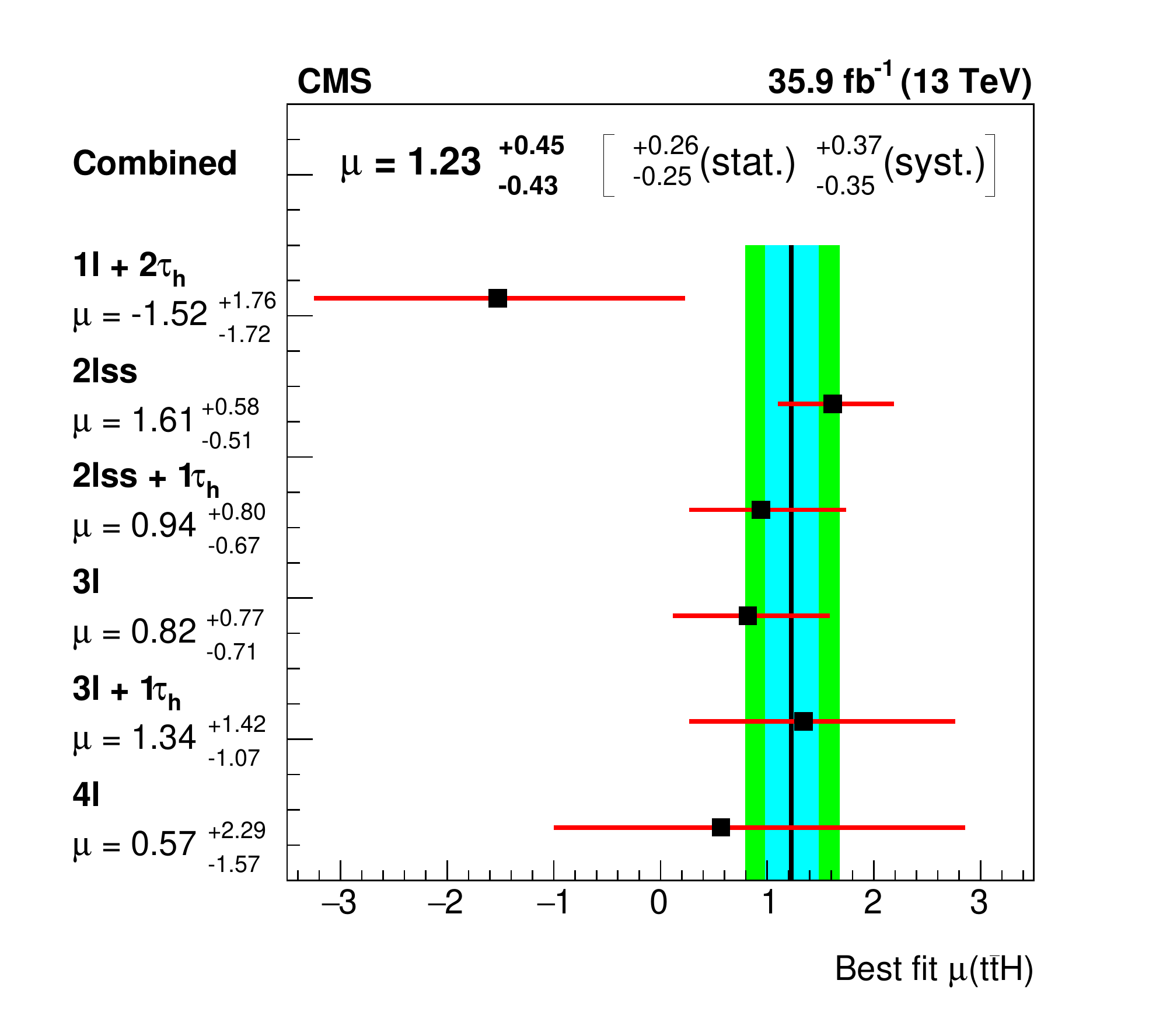}}
\end{minipage}
\hfill
\begin{minipage}{0.5\linewidth}
\centerline{\includegraphics[width=0.9\linewidth]{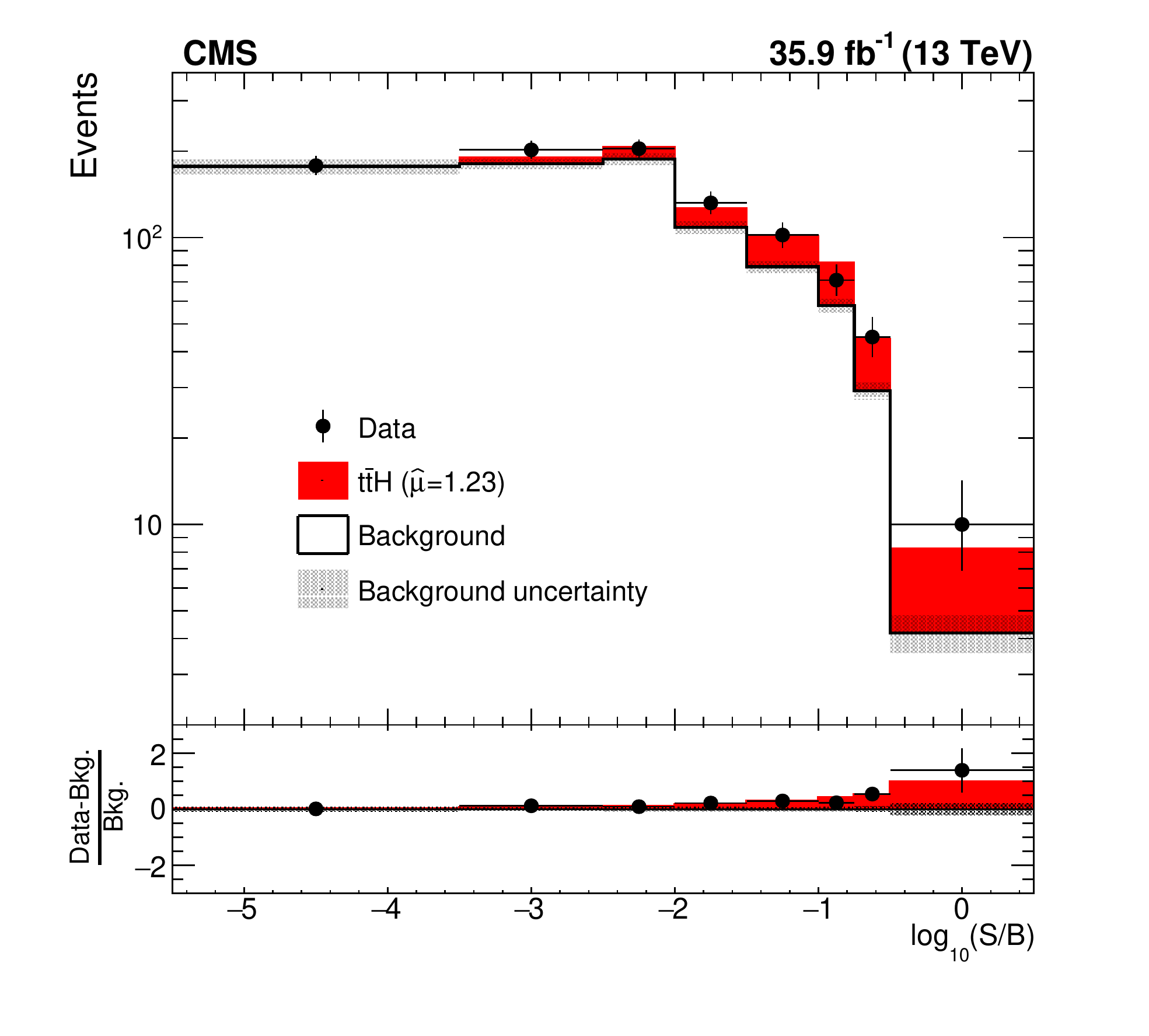}}
\end{minipage}
\caption[]{(Left) Best fit values of $\mu$ measured in each of the individual categories and for the combination of all categories of the multilepton final state~\cite{CMS-HIG-17-018}. (Right) Distribution of the logarithm of the ratio between the expected signal and background in each bin of the distributions used for signal extraction~\cite{CMS-HIG-17-018}. The distributions of signal and background processes are shown for the values of nuisance parameters obtained from the combined ML fit and  $\hat{\mu}$ = 1.23, corresponding to the best-fit $\mu$ value from the ML fit.}
\label{fig:tth_lepton_results}
\end{figure}

The signal is extracted with a simultaneous ML fit to the distribution of final discriminants in all categories. The best fit value of $\mu$ along with its uncertainties are shown in Fig.~\ref{fig:tth_lepton_results} (left) for the individual category and their combination; they agree with the SM expectation. The distribution of log(S/B) from the pre-fit distributions of final discriminants is shown in Fig.~\ref{fig:tth_lepton_results} (right), where the excess of events seen in data is compatible with the expected SM signal. 
The observed significance is 3.2$\sigma$ against an expected significance of 2.8$\sigma$. 

\section{Combination of ${\ttbar}\cPH$ search results}
\begin{figure}
\begin{minipage}{0.5\linewidth}
\centerline{\includegraphics[width=0.9\linewidth]{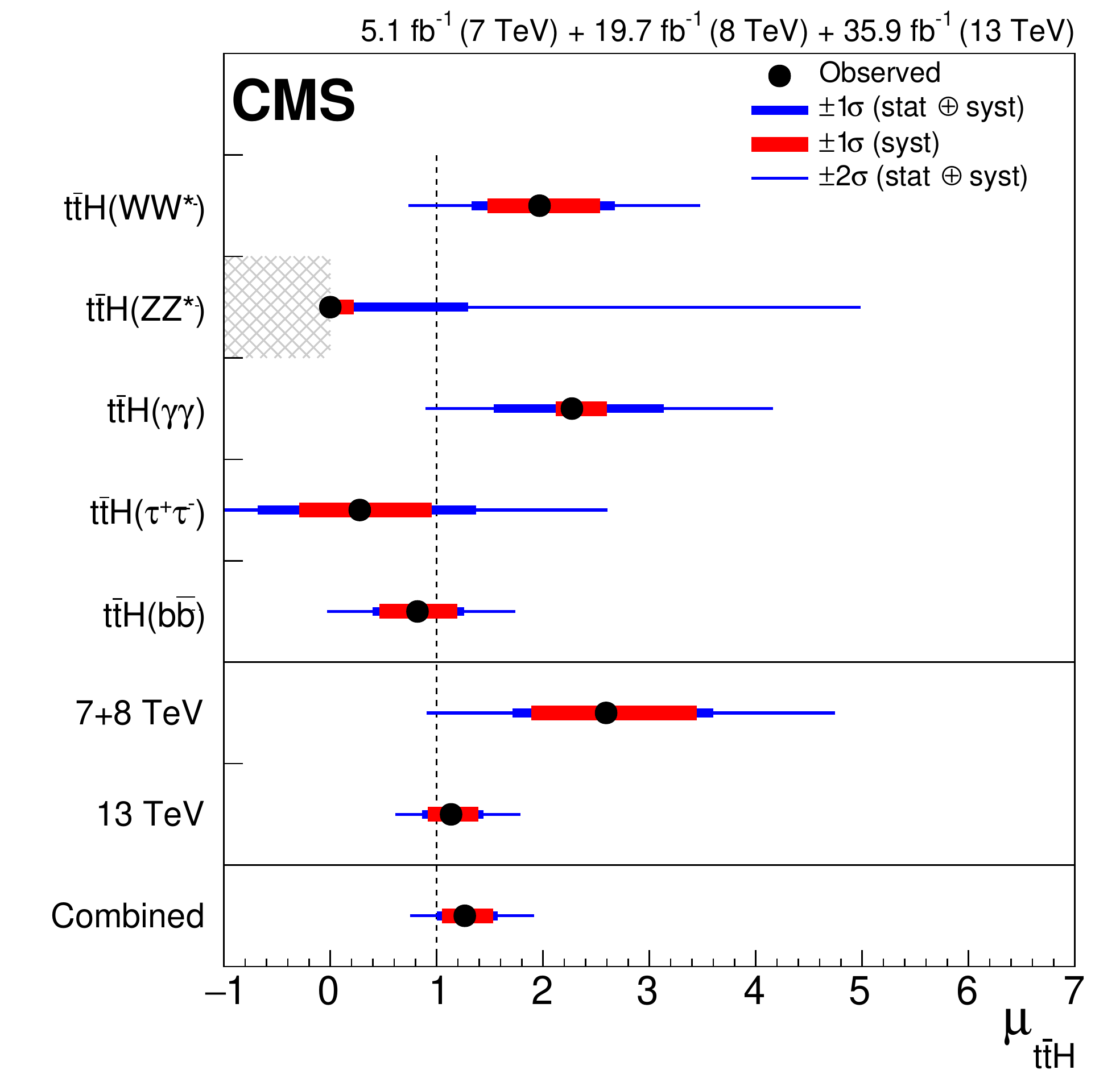}}
\end{minipage}
\hfill
\begin{minipage}{0.5\linewidth}
\centerline{\includegraphics[width=0.9\linewidth]{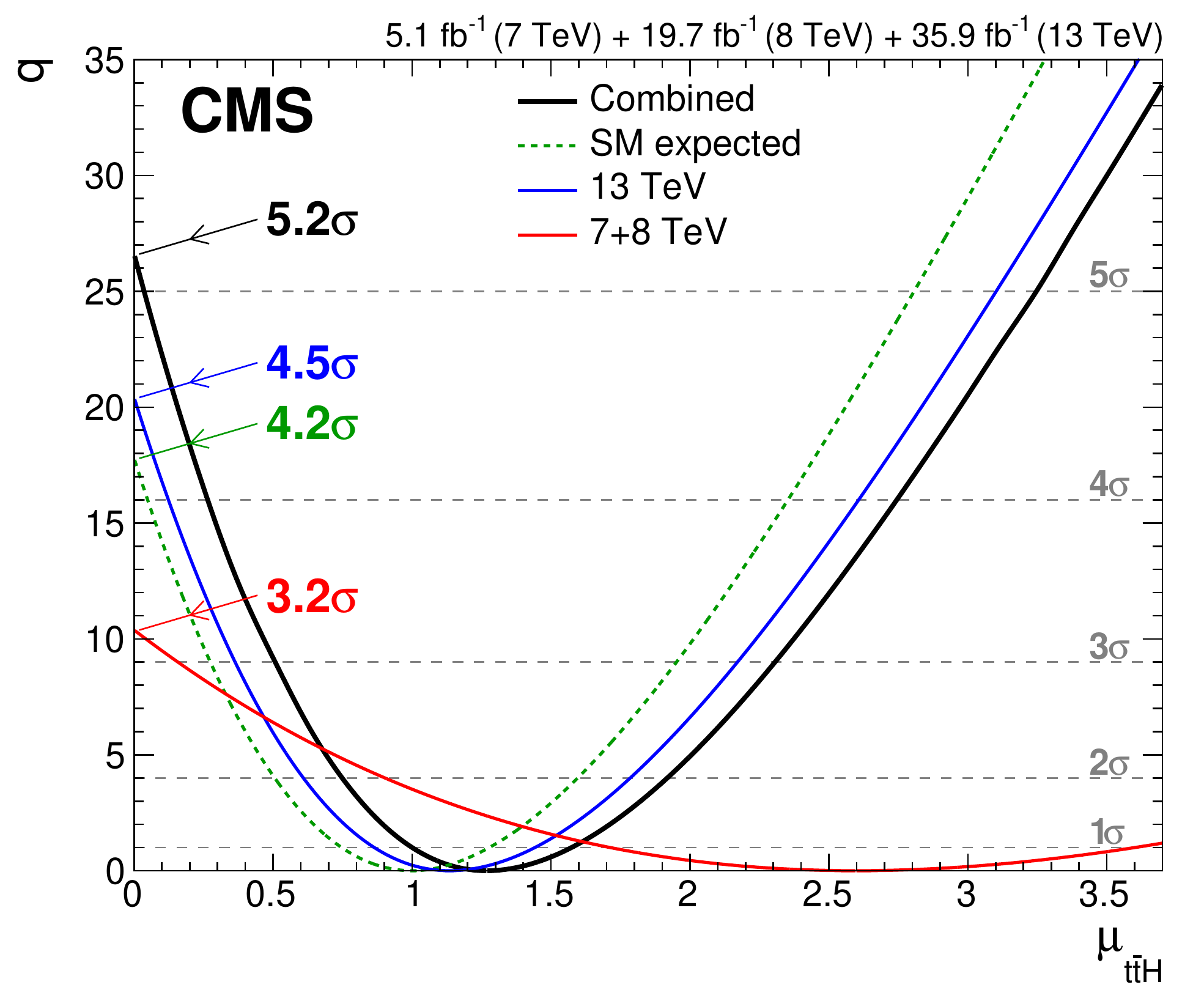}}
\end{minipage}
\caption[]{(Left) Best fit values of signal strength for the five individual decay channels as well as for the combined result~\cite{CMS-HIG-17-035}. (Right) The test statistic as function of signal strength for all decay modes separately for 7+8 $\TeV$ and 13 $\TeV$ and their combinations~\cite{CMS-HIG-17-035}. The expected SM result for the overall combination is also shown.}
\label{fig:tth_combination}
\end{figure}

Results of the searches in different final states, and at the CM energies of 7, 8, and 13 $\TeV$, are combined together~\cite{CMS-HIG-17-035}. The best-fit values of $\mu$ for individual final states as well as their combination are shown in Fig.~\ref{fig:tth_combination} (left). Results are also presented separately for the searches at 7 \& 8 $\TeV$ and at 13 $\TeV$. The measurements for different final states and at different CM energies agree well with each other as well as with the combination. 
Figure~\ref{fig:tth_combination} (right) shows the variation of test statistic ($q$) against $\mu$ separately for searches at 7 \& 8 $\TeV$ and at 13 $\TeV$ as well as for their combination. The $q$ is minimum at the best-fit value of $\mu$. The excess of events observed in data with respect to SM backgrounds is interpreted in terms of significance. The observed significance is found to be 5.2$\sigma$ compared to an expected significance of 4.2$\sigma$. This constitutes the first observation of the ${\ttbar}\cPH$ process at LHC. 

\section{Summary}
Results of the searches for the Higgs boson production in association with a pair of top quarks, based on pp collision data recorded at 13 $\TeV$ by the CMS experiment, are presented. 
These are combined with the earlier results obtained at 7 and 8 $\TeV$ yielding a total significance of 5.2 standard deviations which constitutes the first observation of ${\ttbar}\cPH$ process at the LHC.

\end{document}